\documentclass[sn-mathphys,Numbered]{sn-jnl}
\usepackage{graphicx}
\usepackage{dcolumn}
\usepackage{bm}
\usepackage{mathtools}
\usepackage[english]{babel}
\usepackage{physics}
\usepackage{enumitem}
\usepackage{mathrsfs}
\usepackage{xfrac}
\usepackage{hyperref}
\usepackage{footmisc}
\usepackage{hypcap}
\usepackage{makecell}
\usepackage{url}
\usepackage{ifsym}
\usepackage{bbold}
\usepackage{gensymb}
\usepackage{multirow}
\usepackage{array}
\usepackage{siunitx}
\usepackage{nicefrac}
\usepackage{amsmath}

\jyear{2022}

\title[Analogue gravity and the Hawking effect]{Analogue gravity and the Hawking effect: historical perspective and literature review}

\abstract{
Reasoning by analogies permeates theoretical developments in physics and astrophysics, motivated by the unreachable nature of many phenomena at play. For example, analogies have been used to understand black hole physics, leading to the development of a thermodynamic theory for these objects and the discovery of the Hawking effect. The latter, which results from quantum field theory on black hole space-times, changed the way physicists approached this subject:  what had started as a mere aid to understanding becomes a possible source of evidence via the research programme of `analogue gravity' that builds on analogue models for field effects. Some of these analogue models may and can be realised in the laboratory, allowing experimental tests of field effects. Here, we present a historical perspective on the connection between the Hawking effect and analogue models. We also present a literature review of current research, bringing history and contemporary physics together. We argue that the history of analogue gravity and the Hawking effect is divided into three distinct phases based on how and why analogue models have been used to investigate fields in the vicinity of black holes. Furthermore, we find that modern research signals a transition to a new phase, where the impetus for the use of analogue models has surpassed the problem they were originally designed to solve.
}

\begin{document}

\author*[1]{\fnm{Carla R.} \sur{Almeida}}\email{cralmeida00@gmail.com}
\author*[2]{\fnm{Maxime J.} \sur{Jacquet}}\email{maxime.jacquet@lkb.upmc.fr}

\affil[1]{\orgdiv{Institute of Physics}, \orgname{ University of S\~{a}o Paulo}, \orgaddress{Rua do Mat\~{a}o 1371, Cidade Universit\'{a}ria, \city{S\~{a}o Paulo}, \postcode{05508-090}, \country{Brazil}}}

\affil[2]{\orgdiv{Laboratoire Kastler Brossel}, \orgname{ Sorbonne Universit\'{e}, CNRS, ENS-Universit\'{e} PSL, Coll\`{e}ge de France}, \orgaddress{\street{4 Place Jussieu}, \city{Paris}, \postcode{75005}, \country{France}}}

\maketitle
\section{Introduction}

Because astrophysics is a science made from a distance, it often relies on speculative methods of investigation. For most of history, knowledge about our universe and its constituents was built upon the observation of stars and other celestial bodies, which is made possible only with optical instruments. This restriction gives astrophysics and other large-scale-structure research lines room for a more philosophical approach to the research. In this context, analogies have proven to be an asset to study astrophysical theories. 

The development of black-hole thermodynamics is an interesting case to assess how analogies and effective theories can impact scientific research. With the rise of relativistic astrophysics at the beginning of the 1960s, collapsed objects were considered a plausible explanation for the origin of the then-recently-discovered quasar \cite{greenstein_1964}. In the next decade, the development of a thermodynamic theory of black holes changed how scientists approached the subject. Black holes went from plausible to probable. The scepticism gave place to a general enthusiasm. Identifying black holes as thermodynamical entities in the 1970s \cite{almeida_thermodynamics_2021} allowed for specific predictions that could, in principle, be obtained in the laboratory. In particular, Stephen Hawking's prediction that black holes could radiate~\cite{hawking_black_1974} opened the possibility of reproducing such an effect on Earth through analogue systems. However, analogies had been considered before to investigate black-hole physics. Even the connection between black-hole physics and thermodynamics was, \textit{a priori}, considered an analogy. It helped improve the astrophysical theory but was not considered to be factual. The subsequent identification of black holes as black bodies prompted the formulation of new analogies, this time proposed in support of the theory. In particular, the acoustic analogue of Hawking radiation proposed by William Unruh in 1981~\cite{unruh_experimental_1981} had the ambition of providing evidence to its astrophysical counterpart.

Although nowadays analogue experiments are widely considered an important tool to study inaccessible domains \cite{dardashti_hawking_2019}, the consensus on their limitations and interpretations varies \cite{crowther_what_2018,evans_limits_2020}. Historically, analogies have played different roles at different stages of the development of our current understanding of black holes. Brief reviews of this history have appeared in \cite{barcelo_analogue_2005} and \cite{Field_2021}, each with a different outline for the historical periods about analogue gravity research in the case of black holes. Barceló \textit{et al.} argued that ``analogue models can be reasonably neatly (but superficially) divided into a `historical' period (essentially pre-1981) and a `modern' period (essentially post-1981)'' \cite[p.~31]{barcelo_analogue_2005}. Meanwhile, Fields considered another approach, with three timeline divisions, two of which received the names ``the theory era'' (1981-2008) and ``the experimental era'' (2008-2017) \cite[p.~7]{Field_2021}. These historical outlines by Barceló \cite{barcelo_analogue_2005} and Fields \cite{Field_2021} were pragmatic. Barceló and collaborators underlined the relevance of Unruh's analogue model to the history of analogue gravity, using it as a water divider. On the other hand, Fields added a layer to this view by considering the duality between theory and experiments, proposing a historical division to investigate how analogue experiments can be used as exploratory tools. 

We offer a different perspective, adding to the plural interpretation of this case study. Taking from the philosophical questioning on the motivations for the use of analogies in science, the case of black holes presents itself as more nuanced than previously discussed. In this paper, we address this complexity by proposing a different outline based on epistemological reasoning, investigating the methods and the interpretation of the results to differentiate the objectives for proposing each analogy. We argue that the use of analogue gravity models to assess the physics of black holes can be separated into distinct phases according to the motivation for its use. We identify three historical periods: the \textit{Formative Years}, \textit{the Testimonial Years}, and the \textit{Empirical Years}. The labels we used are emblematic of each phase. Although a single word fails to encapsulate all historical nuances, the labels are useful to identify and communicate the main characteristic of the phase: the Formative Years saw the use of analogies to build upon the theory of black holes, borrowing reasoning from other systems and applying it to this specific astrophysical model. Analogies, in this case, helped to visualise hidden features, thus improving the theory despite their expected fallibility. As for the Testimonial Years, the proposal of Hawking radiation imparted a new purpose for the use of analogies, which aimed at providing testimonials for the highly theoretical conjecture that black holes should radiate. Finally, the Empirical Years mark the era dominated by experimental efforts to observe the Hawking effect in the laboratory. This paper will review each of these phases and explore the transitions between them in terms of their role in analogical reasoning, the aim of the research efforts and the understanding drawn from theory and experiments.

From the presumption that doing history for physics requires integration between history and philosophy of science, we present this paper as a historical case study with a bold move of analysing current research under the same premises. In an attempt to avoid common pitfalls when interpreting case studies, we consider Chang's conceptual moves to avert dilemmas such as historical data manipulation and unreasonable generalisations \cite{Chang.2012}. Also, to sustain the selection of the particular historical cases presented here, we followed the advice of Scholl and R\"{a}z \cite{Scholl.2016}, setting strong criteria for our choices. That is, starting from the conceptual question about the use of analogy theory as a scientific tool, we will present concrete historical cases to instantiate the application of analogies in the astrophysical scenario. 

In the end, we reflect on the current state of the research on analogue gravity, in particular on the novel set of experiments aimed not at observing the Hawking effect but other field effects. Fields found that this new branch of experiments started in 2017 with the first report on the observation of rotational superradiance in a rotating black hole geometry and argued that this is a continuation of research on the `traditional' line of analogue gravity~\cite{Field_2021}. However, in reviewing the literature, we find that although the experiments in question are more recent than those on the Hawking effect, related theoretical studies date back to the early 2000s, just like those on the Hawking effect.
To complete our assessment, we will propose that this new branch evolves separately from the search for the Hawking effect, implying the emergence of a new goal in analogue gravity. Thus, the building of analogue experiments acquires a different status than the one observed in the Empirical phase. To account for this parallel development, we refer to these new dynamics as the \textit{Autonomous Phase}.

This paper is organised as follows: the historical analysis of the first two research phases --- the Formative Years (pre-1974) and the Testimonial Years (1974-2008) --- is presented in Sections \ref{sec:formative} and \ref{sec:testimonal}, respectively. The Empirical Years (2008-2017) are discussed through a literature review in Section \ref{sec:criteria}, connecting history and physics. Section \ref{sec:autonomous} brings a reflection on the autonomy of the analogue-gravity research programme.

\section{The Formative Years: pre-1974}\label{sec:formative}

After the renaissance of general relativity during the mid-fifties \cite{blum_renaissance_2016}, the theory about gravitational collapse quickly advanced, with new solutions discovered \cite{kerr_gravitational_1963,newman_metric_1965} and several features unveiled, such as the presence of an accretion disk around spinning black holes \cite{zeldovich_fate_1964}, the theorems on the existence of singularities \cite{hawking_singularities_1970}, and a mechanism to extract energy from rotating black holes \cite{penrose_extraction_1971}. In the sixties, astronomers identified quasars, quasi-stellar radio sources, as possible candidates to be black holes \cite{greenstein_1964}, but this identification was speculative. Most notably, the term \textit{black hole} became popular after John Archibald Wheeler adopted it as a technical term in 1967, when before it was used metaphorically \cite{herdeiro.2018}. As exemplified, the sixties and early seventies were formative years for the theory of black holes.

At that point, a black hole was a classical object --- its physics wholly described by general relativity. It was a highly theoretical entity, so it is no surprise that analogies appeared in an attempt to understand or build upon the black-hole gravitational model. Although few, the analogies proposed had one thing in common, they all compared black holes to widely different systems in which no actual connection was expected. A cylindrical resonator, a fish falling in a waterfall, a black body --- those identifications had no physical meaning at that point, but they guided theoretical research on the source system. The analogies helped increase visualisation or understanding of the black-hole theory. Nevertheless, they could have led to false conclusions, as the comparisons were not considered factual.

To prove our point, we will discuss next the first analogies that emerged in this astrophysical context: Zel'dovich's analogue model, Unruh's fish-in-the-waterfall visualisation tool, and the thermodynamical analogy to black-hole mechanics. We end the section with a comment on the transition to the next stage in the history of analogue gravity.

\subsection{Zel'dovich's analogue model}
\label{subsec:Zeldovich}

Yakob B. Zel’dovich learned about Roger Penrose’s mechanism to extract energy from a rotating black hole from ``stimulating discussions'' with Charles Misner, Kip Thorne, and John Wheeler. The Penrose Process, as it became known, refers to a particle near the event horizon of a Kerr black hole being accelerated to the infinite after stealing the black hole’s rotational energy and slowing down its spin in the process. Zel’dovich, a versatile physicist with a background in nuclear physics, tried to understand Penrose’s mechanism idea by comparing a black hole to an entirely different system. The acceleration of the particle to infinite in a Kerr black hole would work similarly to the amplification of scattered incident waves in a resonator.

In his 1971 paper \cite{zeldovich_generation_1971} published in the Soviet journal \textit{ZhETF Pisma Redaktsiiu}, Zel'dovich considered an axially-symmetrical body rotating inside a resonator cavity. He showed that the resonator operates as an amplifier for waves with particular wavelengths in an absorbing medium at rest for a given coordinate system. In the process of scattering incident waves, the waves with momentum parallel to the rotation would be amplified. Zel'dovich noticed that the Planck constant disappears in the expression for the condition for amplification of waves --- ``a gift to the modern method of expression in an era when `quantum mechanics helps understand classical mechanics'\,'' \cite[p.~181]{zeldovich_generation_1971}. This scattering phenomenon in this resonator system had not been predicted before. In his attempt to understand the classical gravitational mechanism called Penrose Process, Zel'dovich predicted a new phenomenon in an analogue quantum system.

Moreover, in the presence of the resonator, the scattering should lead to generation. A rotating body made of a material that absorbs waves when at rest would generate electromagnetic radiation due to interaction with vacuum fluctuations. Zel'dovich added that ``apparently,'' a similar situation would happen when considering a rotating body in the state of relativistic gravitational collapse. However, the effect would be negligibly small, Zel'dovich concluded. 

In December of the same year, Zel’dovich submitted an extended version \cite{zeldovich_amplification_1972} of this paper, in which he focused on the amplification of waves phenomenon. The first paper had a more casual approach to the problem. Thus, Zel’dovich worked on a more thorough proof. An infinite cylinder of low conductivity would amplify reflected electromagnetic waves of varying polarisation. Future Nobel-laureate Pyor L. Kaptiza drew Zel’dovich’s attention to the fact that this situation would be ``analogous to the amplification of sound by reflection from a resting-medium boundary that moves with supersonic velocity'' \cite[p.~1085]{zeldovich_amplification_1972}, apparently previously described by N.N. Andreev and I.G. Rusakov in 1934. This second analogy was mentioned briefly and did not have a purpose in Zel’dovich’s reasoning.

Zel’dovich expanded on the other analogy: the comparison between his model and the astrophysical scenario. Addressing the similarities between the resonator and gravitational systems, Zel’dovich remarked once again that generation of waves may occur in the vicinity of a Kerr black hole and credited the phenomenon, ``if it exists,'' to a supposed instability of the Kerr metric \cite[p.~1086]{zeldovich_amplification_1972}. Although Zel'dovich implied that an analogous phenomenon would happen in the astrophysical scenario, the hesitation about properly predicting the spontaneous generation of energy in the case of black holes is noticeable. The analogue phenomenon he predicted, having been just proposed, had not yet been detected. Furthermore, the translation between the quantum field theory applied to his model to the classical general relativity describing the vicinity of a black hole could be flawed. The spontaneous generation prediction was taken with a grain of salt. Maybe it was a possibility to be explored and not a genuine effect.

Nevertheless, Zel’dovich’s analogue model indeed helped to advance the theory of black holes. Alexei Starobinskii, Zel’dovich’s postdoctoral student at the time, ditched the analogy and went straight to the original problem. He followed the lead of his advisor and published a purely astrophysical version of Zel’dovich’s papers, using quantum field theory in the neighbourhood of a Kerr black hole to reinterpret the Penrose Process as a semi-classical process \cite{starobinskii_amplification_nodate}. He did not, however, mention the generation of waves in a Kerr metric in the paper. Starobinskii presented his and his supervisor's results to Stephen Hawking, who would later lead the research in a new direction.

\subsection{A fish in a waterfall}\label{subsec:waterfall}

The physics of black holes is highly counter-intuitive. Thus, over the decades, attempts to understand what these highly theoretical objects are and how gravity acts around them often relied on metaphors and analogies to improve visualisation.\footnote{ For more on the use of visualisation as tools for understanding, we refer to \cite{de_regt_visualization_2014}. } One clever analogy compared a transmitter crossing the event horizon of a black hole to a fish falling in a waterfall.

\begin{figure}[ht]
    \centering
    \includegraphics[scale=0.5]{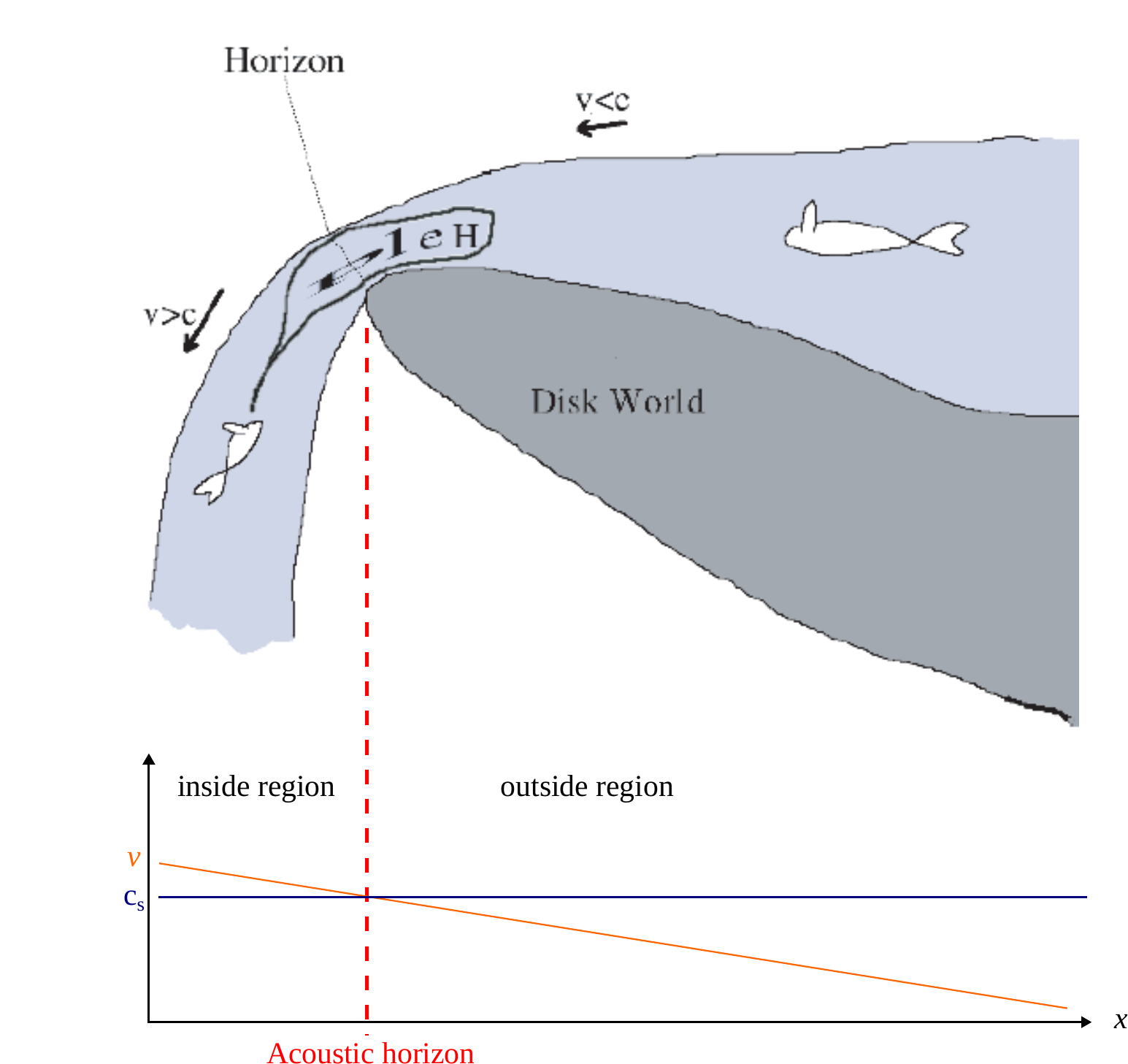}
    \caption{\textbf{The river model of space-time}. \textbf{Top: Unruh's illustration} presented in \cite{unruh_analogue_2007} (reproduction with the author's permission). Two fish in the fantasy world of Discworld. The falling fish calls for help, but the \textit{p} is emitted after it crosses the horizon, stretching out without ever reaching the other fish.
    \textbf{Bottom: transsonic fluid flow forming an acoustic horizon}. The fluid flow velocity $v$ (orange) increases from right to left, exceeding the speed of sound $c_s$ (blue) at the acoustic horizon~\cite{unruh_experimental_1981}. This surface separates two regions, the outside, where the flow velocity is sub-sonic, and the inside, where the flow velocity is super-sonic. The speed of sound may also be spatially dependant~\cite{visser_acoustic_1998}.}
    \label{fig:fish}
\end{figure}

William Unruh was a Ph.D. candidate at Princeton and finished his thesis about Dirac particles and geometrodynamical charge in curved spaces in 1971 under the supervision of John A. Wheeler. When tasked with the job of explaining black-hole physics at a colloquium, Unruh came up with a simple analogy that would help his audience to understand how black holes trap light. In his recollections, Unruh often mentions this colloquium being at Oxford University in 1972. Unruh compared a fish falling into a waterfall (Figure \ref{fig:fish}) in which the velocity of the water is greater than the speed of sound to the light-trapping mechanism inside the event horizon of a black hole. In this scenario, any noise made by the fish would not reach the top of the waterfall. Similarly, the light signal sent from inside a black hole would not reach the horizon \cite[p.~3]{unruh_analogue_2007}.\footnote{It is important to underline that this information is given through recollections. In the paper in which Unruh recalled this episode \cite{unruh_analogue_2007}, his account of the analogy is more refined and poetic in a way it could not have been proposed in 1972. For example, instead of the waterfall, he used the fantasy world of Rimfall, created by author Terry Pratchett in his fantasy series Discworld. The first book in the Discworld series was published in 1983, after Unruh's 1981 paper proposing the formal analogy. Nevertheless, it is plausible and likely that a simpler version had been proposed earlier, around the year Unruh recalled.} At the time, this analogy had pedagogical purposes, an attempt to improve visualisation of the phenomenon.

Notice that this analogy differs from Kaptiza's observation about the similarities between sound amplification and Zel'dovich's resonator experiment mentioned in Subsection \ref{subsec:Zeldovich}. Kaptiza's comment was meant to support the idea Zel'dovich had, and it was one step away from the case of black holes. It was a remark about the low-conductivity, infinite cylinder instead. 

Unruh's fish analogy was a simpler comparison to help his audience understand the properties of the event horizon at a time when black holes were not well known. Unruh recalled reading Zel'dovich's paper at some point but not Kaptiza's fluid analogy.\footnote{Private communication.} It is possible, however, that it was a source of inspiration for his fish analogy, even if unconsciously. Since it did not have research purposes, it is clear the analogies were expected to be incompatible on a theoretical level with the source system. The fish-in-the-waterfall analogy could have been forgotten if Unruh had not revisited it later in a formal proposal for a sound-analogue model.

\subsection{The thermodynamical analogy}\label{subec:thermo}

Jacob Bekenstein proposed the formulation of a thermodynamic theory of black holes in his Ph.D. thesis \cite{bekenstein_baryon_1972} and later in his paper \textit{Black Holes and Entropy} \cite{bekenstein_black_1973}. Comparing black-hole physics and the laws of thermodynamics, he proposed that black holes had actual entropy and thus were thermodynamical entities.\footnote{For a deeper analysis of the historical events that led to the development of the thermodynamical theory of black holes, see~\cite{almeida_thermodynamics_2021}.} At the time, the comparison drawn by Bekenstein was appealing, but not his conclusions. Actual entropy implied actual temperature, but no thermal content could escape from within the horizon. Bekenstein had suggested the identification of the event horizon's area to entropy using information theory, but he concluded that the analogue temperature was effective. This inconsistency led to criticism and disbelief, although the idea survived as an analogy.

John M. Bardeen, Brandon Carter, and Stephen Hawking explored the analogue idea further, publishing the four laws of black-hole mechanics~\cite{bardeen_four_1973} derived from the comparison to the four laws of thermodynamics. The second law relied on the analogy between the area $A$ of the event horizon and the system's entropy. A black hole's area (entropy) does not decrease with time. The first law established an analogy between the surface gravity $\kappa$ and temperature, similar to the comparison between the event horizon's area and entropy. The zeroth law says the surface gravity (temperature) is constant over the event horizon, and finally, the third law, left unproven \textcolor{red}{by the authors},\footnote{A proof of this law was provided by Werner Israel in 1986~\cite{Israel_1986}, over a decade after Badeen, Carter, and Hawking's formulation.} states that ``it is impossible by any procedure, no matter how idealised, to reduce $\kappa$ to zero by a finite sequence of operations''~\cite[p.~169]{bardeen_four_1973} Considering the correlation between $A$ and entropy and $\kappa$ and temperature, this law is equivalent to the third law of thermodynamics, \textit{A system's entropy approaches a constant value as its temperature approaches absolute zero.}

As an analogy, the comparison between thermodynamics and black-hole physics was fruitful. For example, Bardeen, Carter, and Hawking argued that, while this third law remained open, there were ``strong reasons for believing in it.'' \cite[p.~169]{bardeen_four_1973} One of the arguments suggests that, if this law was untrue, a naked singularity could exist in violation of the cosmic censorship conjecture. Despite this, the analogy led to not-acceptable conclusions. Identifying the area as entropy and surface gravity, that is, the gravitational acceleration that would be experienced at the event horizon, as the temperature would lead to undesirable outcomes, such as that black holes could emit radiation.

The thermodynamic analogy furthered the theory but presented ingrained flaws, namely the disparity between the actual temperature and the analogous one. Although Bekenstein claimed from the beginning that the comparison was more than an analogy, he remained alone in his understanding of black holes as thermodynamical objects. 

\subsection{The transition to the testimonial years}

The three analogies presented here are of different natures. Zel'dovich proposed an actual experiment that would act similarly to a black hole. Unruh's analogy was an artistic image created for pedagogical reasons, an illustration to provide insight into the gravity near a black hole. Meanwhile, the comparison between black-hole physics and thermodynamics was theoretical. However, the three analogies shared a purpose: to drive forward the theory of black holes or improve its visualisation. They also present incongruous conclusions, considering what was known at the time. Unruh's imagery was never meant to be factual since gravity was not expected to behave as a fluid near a black hole. Zel'dovich's analogy suggested that rotating black holes could radiate, and the thermodynamic analogy suggested the black hole would have a non-zero temperature, predictions incompatible with the fact that nothing could escape the event horizon. 

The situation changed when Hawking predicted that all black holes should radiate in 1974 \cite{hawking_black_1974}. Hawking's calculations showed that the radiation rate implied that the temperature of a black hole would be precisely what was obtained in Bekenstein's analogy. It was the missing piece of the puzzle. With the theory providing a physical mechanism to define temperature, the identification of black holes as thermodynamical objects were uncomplicated, despite the \textit{ad hoc} merge of classical and quantum theory in Hawking's paper. The analogy with thermodynamics had achieved its goal: it had advanced the theory and led to a new interpretation of black holes. Eventually, it lost the status of analogy altogether, and a new phase of investigation began.

\section{The Testimonial Years: 1974-2008}\label{sec:testimonal}

As for what concerns us, the theoretical discovery of the Hawking radiation established the understanding that black holes were thermodynamical objects, so the comparison between black hole physics and thermodynamics~\cite{bekenstein_baryon_1972,bekenstein_black_1973,bardeen_four_1973,almeida_thermodynamics_2021} was considered to be more than a mere analogy~\cite{jacobson_black-hole_1991,unruh_notes_1976}. Hawking reflected on the situation at the end of his 1974 paper. ``Bekenstein suggested on thermodynamic grounds that some multiple of $\kappa$ should be regarded as the temperature of a black hole. He did not, however, suggest that a black hole could emit particles as well as absorb them. For this reason, Bardeen, Carter and I considered that the thermodynamical similarity between $\kappa$ and the temperature was only an analogy. The present result seems to indicate, however, that there may be more to it than this.'' \cite[p.~31]{hawking_black_1974}

With this breakthrough, new analogies appeared. Before, the comparison between the classical gravitational theory and other systems had been based on symmetry in the behaviour of said theories, even though their nature differed. However, Hawking's prediction changed this perception. Now, black holes were described not solely by the classical gravitational theory but also by thermodynamics and quantum field theory. Thus, studying the astrophysical scenario through analogies became more than a mere comparison to understand the theory. The reproduction of quantum effects similar to Hawking radiation in different systems acquired the status of indirect evidence for its astrophysical counterpart. The use of analogies aimed at cementing Hawking's results, given that observational evidence was not attainable at that point.\footnote{It remains unattainable, for that matter.} 

To support this view, we will first explain Hawking radiation as it was first idealised and then discuss the various reinterpretations of this phenomenon. At this point, it becomes clear that analogies were not the sole way to provide testimony for the existence of Hawking radiation. The first reinterpretation of the phenomenon came through an effective theory, where gravity was emulated by accelerated particles. The discussion on the realism of effective theories revolves around its limitations \cite{rivat.2021}, much like the case for analogue systems. From an epistemological point of view, the two --- analogue models and effective theories --- served the same purpose, helping us to identify the objectives of using analogies in this context.

\subsection{The Hawking radiation and black hole evaporation}
\label{subsec:bhevapo}

``Quantum gravitational effects are usually ignored in calculations of the formation and evolution of black holes,'' Hawking explained \cite[p.~30]{hawking_black_1974}. Following Starobinskii's lead, Hawking used quantum field theory to uncover hidden features of back holes that were the root of otherwise paradoxical results when considering Bekenstein's thermodynamical analogy. Hawking identified an imbalance in the creation/annihilation process, resulting in more particles being created and emitted to infinity. In brief, the extreme curvature of space-time near the horizon causes positive- and negative-norm modes to scatter, which results in a mixing of the creation and annihilation operators of the field, yielding spontaneous emission from the vacuum.  The rate of emission over absorption Hawking calculated to be $(\exp{2\pi \omega / \kappa} - 1)^{-1}$, the one expected from a body with a temperature of $\kappa / 2\pi$, in geometrical units. That is, an observer in the asymptotic future would measure a thermal spectrum of temperature inversely proportional to the black hole mass.

With this emission rate, Hawking predicted that black holes could eventually evaporate. He had shown that, although the emission rate was negligible locally, in a time scale of the universe's lifetime, the emission could add up to be significant in the mass measurement of the black hole, and it could result in complete evaporation \cite{hawking_particle_1975}. It is worth noticing that Hawking radiation is not a classical phenomenon, at least not in its original presentation.

Hawking's prediction has raised numerous questions ever since, most of which remain unanswered to date: where are the particles emitted from (see, e.g.~\cite{unruh_origin_1977,jacobson_black-hole_1991}), what is the source of this radiation (see, e.g.~\cite{hawking_particle_1975,corley_hawking_1996}), how does this process obey causality and conservation of energy (see, e.g.~\cite{davies_energy-momentum_1976}), what would a freely falling observer see as they approach and cross the event horizon (see, e.g.~\cite{davies_thermodynamics_1978}), the influence of the gravitational potential on the emission spectrum (see, e.g.~\cite[p.~260]{birrell_quantum_1994}), how does Hawking radiation fit in the picture of black hole information (see, e.g. \cite{skenderis.2008, ashtekar.2020, raju.2022}), and so on. 

However, historically, the Hawking radiation changed a paradigm concerning the interpretation of analogies. This prediction validated a (once considered to be an) analogy, and, in the process, it reinforced the argument in favour of using analogical reasoning to assess theoretical works. 

\subsection{Gravity emulated by accelerated particles}\label{subsec:mirror}

Before Hawking had predicted that black holes should radiate, Stephen Fulling, an American mathematician and physicist, was puzzled by the nonuniqueness of canonical field quantization in Riemannian space-time \cite{fulling.1973}. He had concluded that the interpretation of particles would change from the standard one in a special-relativity free-field theory, that is, in flat space-time with an accelerated coordinate frame, the \textit{Rindler space}. This space has a horizon with similar properties to the static exterior region of a Schwarzschild black hole. Fulling found that quantising a Rindler space can be interpreted as the physical situation of a reflective wall located at the light cone. Comparing the Rindler space to the Schwarzschild geometry, he noted that ``if this interpretation is correct, field quantization in the exterior Schwarzschild solution regarded as a static space-time corresponds to the presence of a reflecting wall at the Schwarzschild radius.'' \cite[p.~7]{fulling.1973} That is, particles could be scattered by the event horizon. 

The results were odd, and Fulling constantly raised questions about its ontological reality, arguing that the situation was not well-understood enough. Curiously, his objections to the conclusions about particle physics came from the comparison to a Schwarzschild black-hole system and the previous knowledge that the event horizon is a one-way-in membrane. Fulling used black holes as an asset to assess a field-particle theory.

Hawking's paper presenting the black-hole radiation was published the following year. The prediction that there would be an imbalance between particles created and annihilated near the horizon of a black hole was surprising, to put it in the words of British physicist Paul Davies. He and J.D. Taylor had previously assumed, in a brief communication published in Nature in 1974, that ``there is not creation of massless particles in the exterior region of a Schwarzschild black hole,'' \cite[p.~37]{davies.1974} although they concluded it could happen for a Kerr (rotating) black hole --- essentially Zel'dovich's prediction. Intrigued by Hawking's work, Davies revised this previous assumption in a full paper published in the next year \cite{davies.1975}. Davies revisited Fulling's paper and transposed the black hole problem to a Rindler system to verify if replicating Hawking's steps would also lead to radiation. He justified his approach, ``An analysis of flat space quantum field theory in Rindler coordinates may therefore be regarded as a test case for the Schwarzschild system, with the added advantages of conceptual simplicity and the knowledge that the `correct' quantization scheme based on Minkowski coordinates is always available for comparison.'' \cite[p.~610]{davies.1975} 

He concluded that particle production is expected indeed but was too puzzled by the problem raised by Fulling. The nonuniqueness in the definition of the concept of particles in the black hole case implied that different observers would detect a different rate of particle creation. He mentioned that ``[i]ndeed Hawking has proposed that a \textit{freely-falling} observer would not see a large amount of particle production near a black hole as he approached the horizon. However, Hawking advances the comparability of long wavelengths with horizon size as evidence for low-frequency particle production, rather than for a break down in the definition of particle in this r\'{e}gime.'' \cite[p.~615]{davies.1975} In the end, Davies expressed an acceptance of the idea of black hole radiation. Concerning Fulling's particle-definition paradox, he remarked that Hawking's work could reveal plenty about both black holes and elementary particle physics. It is interesting to note that while Fulling used black holes to raise questions about his own conclusions, Davies used Fulling's work to corroborate Hawking's results.

The issue about particle definition raised by Davies was addressed a year later by William Unruh, who also used a Rindler-Minkowski system to study the scattering effect. As previously mentioned, William Unruh’s interest in black holes can be traced back to the beginning of the seventies, but his first publication on the subject came in 1976. Unruh debuted his research on the topic with a paper titled \textit{Notes on Black Hole Evaporation}, where he used boundary conditions to emulate the horizon, and acceleration on flat space to emulate gravity in a two-dimensional model, an approximation for a spherical symmetric four-dimensional black hole. His intentions were clear: to investigate Hawking’s suggestions that black holes behave as an almost ideal black body with an effective temperature of $(8\pi M)^{-1}$, where $M$ is the mass of the black hole \cite[p.~14]{unruh_notes_1976}.

Unruh’s paper was more detailed than Davies’s \cite{davies.1975}, also addressing the problem with the definition of particles by considering the Rindler coordinates as parametrisation for a detector in an accelerated motion. Hawking had predicted that a detector moving along a geodesic would not detect the radiation. In Unruh’s view, this problem was connected to the challenge of defining what a vacuum state is in a gravitational field. ``The key question is: which of these definitions, if any, corresponds to what the gravitational field sees as particles, i.e., as carriers of energy and momentum? If one accepts the ‘geodesic’ particles, both near the horizon and at infinity, as the true particles, one obtains the paradoxical situation that particles both flow into the horizon and out through infinity --- i.e., the `black hole' increases in area, and loses mass to infinity. The only place where the `energy' for such a process could originate would be in the vacuum polarization of the field in the regions outside the black hole.'' \cite[p.~888]{unruh_notes_1976} 

Unruh concluded that what an inertial observer would call a vacuum would, in fact, contain many particles in thermal equilibrium for an accelerated observer. The novelty in Unruh's proposition compared to Fulling's was the thermodynamical interpretation of the change from resting to accelerated frames, when Fulling had suggested prior a change in the definition of particles. Transposing the problem to the case of black holes, Unruh calculated that a particle detector stationed near the horizon functioned similarly to a static detector in Rindler coordinates. This effect of heat generated by acceleration became known as the \textit{Unruh effect}.

This re-interpretation of a Rindler system as a black hole does not constitute an analogy but an effective theory. It is clear that both Davies and Unruh used the investigation of a Rindler system as a tool to support Hawking's proposal, presenting the results as evidence of it despite the questions it also brought.

\subsection{Black holes in the laboratory?}\label{subsec:unruh}

The phenomenon of amplification of waves had been informally compared to sound moving in a supersonic medium before by Kaptiza in a remark about Zel'dovich's experiment (Subsection \ref{subsec:Zeldovich}), although not in a connection to black-hole physics. Around 1980, when teaching a course on fluid mechanics, Unruh wondered if there was more to the fish analogy he had invented as a pedagogical tool back in 1972 (see Subsection \ref{subsec:waterfall}). Investigating how the sound waves would actually behave in this fish scenario, Unruh found that the equations were equivalent to those of a scalar field in a pseudo-Riemannian metric. 

Comparing Kaptza's and Unruh's arguments we see that Unruh's analogy came from other reasoning and had a different purpose: to find evidence for the Hawking radiation in the laboratory. He said, ``[T]he experimental investigation of the [Hawking radiation] phenomenon would seem to be virtually impossible, and would depend on the highly unlikely discovery of a small black hole (...) near the Earth. // However, a physical system exists which has all of the properties of a black hole as far as the quantum thermal radiation is concerned, but in which all of the basic physics is completely understood. In this system one can investigate the effect of the reaction of the quantum field on its own mode of propagation, one can see what the implications are of the breakdown of the wave equation at small scales on the evaporation process, and one might even contemplate the experimental investigation of the thermal emission process.'' \cite[p.~1351]{unruh_experimental_1981}

Unruh considered the motion of sound waves in a convergent, irrotational fluid flow. He found that the equation of motion for this massless scalar field was the same as for the motion of a scalar wave in a geometry with the metric,\footnote{We chose to use the notation Unruh used in his 1995 paper \cite{unruh_sonic_1995} instead of the original for simplicity since our goal with this exposition is a better understanding of the analogy.}
\begin{equation}
\label{eq:unruh_metric_1}
    ds^{2} = \rho \left( c^{2} - v^{2} \right) d\tau^{2} - \left( \delta_{ij} + \frac{v^{i} v^{j}}{\left( c^{2} - v^{2} \right)}\right) dx^{i}dx^{j} \,.
\end{equation}
where $c^{2}=dp/d\rho$ is the local velocity of sound. Considering $c^{2}$ to be constant, a spherically symmetric background flow, and the existence of a region where the fluid exceeds the velocity of the sound, equation \eqref{eq:unruh_metric_1} becomes,
\begin{equation}
    ds^{2} = \rho \left[ \left(c^{2} - v^{2} \right) d\tau^{2} - \frac{1}{1-\frac{v^{2}}{c^{2}}}dr^{2} - r^{2}d\Omega^{2} \right] \,,
\end{equation}
with $d\Omega = d\theta^{2} + \sin{\theta}^{2} d\phi^{2}$. 

The comparison to the Schwarzschild metric came naturally, and so Unruh called the system a \textit{sonic black hole}. Near the horizon, the sonic black hole would emit sound waves in a thermal spectrum that Unruh calculated to be the temperature of a black hole if the speed of sound were to be replaced by the velocity of light. Unruh remarked that this temperature would be low and thus probably undetectable in the presence of turbulent instabilities. ``It is, however, much simpler experimental task than than creating a $10^{-8}$-cm black hole.'' \cite[p.~1353]{unruh_experimental_1981}.

This sound analogue allowed physicists to dream about experimental evidence for an astrophysical phenomenon. It certainly did not answer the questions that the proposition that black holes should radiate had created, but it at least gave a practical direction for a theoretical prediction.

\subsection{Viable experiments}\label{subsec:viabexper}

Even if philosophically controversial, the use of effective theories and analogies to investigate a theoretical phenomenon that would happen around an astronomical body whose existence had not yet been confirmed gave a solid direction to research. The possibility of replicating some of the black hole's features in a laboratory supported Hawking's theoretical prediction and presented a way to test it, even if indirectly. It made the research tangible. ``All of us agree that healthy theory means beautiful theory and beautiful experiment,'' summarised H. Ro\c{s}u in his proposal titled \textit{Towards Measuring Hawking-like Effect in the Laboratory}, presented to the Central Institute of Physics Astronomy \& Space Research Center in Bucharest, Romania, in 1989 \cite{rosu.1989}\footnote{An updated version of the proposal exists in~\cite{rosu_estimates_1994}.}.

After Unruh's 1981 paper, several propositions of viable experiments appeared. Not all of them came from a direct analogy, like the one proposed by Unruh. At this point, the discussion about the existence of Hawking radiation became intrinsically connected to the existence of the Unruh effect and the theory surrounding his analogue model. The scattering phenomena described in the Rindler space by Fulling, Davies, and Unruh, or in the creation of sound waves near the horizon of a sonic black hole, were treated as the \textit{Hawking-like effect} or \textit{the Hawking effect}. 
The possibility of detecting the Unruh effect with particle accelerators was considered by Bell and Leinaas \cite{bell_1983} in their 1983 paper \textit{Electrons as Accelerated Thermometers}. In 1987, Nugayev proposed to detect particle creation by a black hole in terms of temperature corrections to the Casimir effect \cite{nugayev_1987}, arguing that the flow of negative
Casimir energy would cause the area of the horizon to shrink at a rate consistent with Hawking's prediction, providing another route to demonstrate Hawking-like effects in the laboratory. A detector for observing the Unruh effect by acceleration was proposed in 1988 \cite{rogers_1988}. 

The common sentiment in those propositions was to provide testimony for the astrophysical scenario. This becomes apparent in Theodore Jacobson's 1991 paper, \textit{Black Hole Evaporation: An open question} \cite{jacobson_black-hole_1991}. Learning about the topic from Unruh himself, American physicist Ted Jacobson gave a panoramic overview of the topic at the time. Despite the title, he described the problem not as a question but instead leaning toward an affirmative answer. One of Jacobson's arguments to support the idea that black holes evaporate was the analogy between the black hole environment and Unruh's \textit{fluid-flow analogue}.\footnote{Jacobson's original spelling was `fluid-flow analog.' In the general literature, both the British `analogue' and US `analog' spellings are used.} In papers from this decade onward, we notice a tendency to expand the theory to encapsulate analogue models. The \textit{Hawking effect} became a general (scattering) phenomenon to which the Hawking radiation is only one example. The expectation now was that detecting the Hawking effect in a particular model would imply that it would happen in other systems, including the astrophysical. 

\subsection{The transition to the Empirical Years}

Undoubtedly, the Hawking radiation was an intriguing phenomenon. Zel'dovich had proposed earlier that spontaneous emission could occur in his resonator system, but it had not been detected at that point. The proposition that it would also happen around a black hole was fascinating. It pushed physicists to try to understand this spontaneous emission through analogies and effective theories. However, this time the objective was not to develop the theory but to provide testimony to it by identifying the same phenomenon in different scenarios. If Hawking-like phenomena, the Hawking effect, were to be observed in different systems, could it be empirical evidence for the existence of the astrophysical counterpart around black holes?

The search for evidence of this phenomenon through analogue experiments and effective theories was possible because of the identification of black holes as thermodynamical objects. While gravitational theories remained falsifiable only through observations, thermodynamics and quantum field theory allowed for experimentation. It justified searching for evidence in analogies, even if indirectly. 

For decades, physicists strengthened the theory to build those experiments, but the technological aspect hindered the development of practical proposals. Eventually, the technology caught up. We refer to a single proposition as an era divider here to establish a clear timeline, but the transition was more gradual than this choice suggests. 

\section{The Empirical Years: 2008-2017}
\label{sec:criteria}

During the first decade of the millennium, the investigation on analogue gravity expanded to the wider goal of reproducing the Hawking effect and became a proper research programme, with several groups dedicating their resources to this matter. The connection to the astrophysical scenario remained implied, and the consensus among this community was that the detection of the phenomenon in analogue gravity could be taken as confirmation of the Hawking radiation. In this section, we discuss the growth of the analogue-gravity research field and its implications.

However, instead of case studies, we will briefly review the theoretical framework on analogue gravity in the early 2000s, focusing on a few examples that illustrate two complementary dynamics: the generalisation of the Hawking effect to an increasing number of physical systems and the advent of detailed proposals for its experimental detection. Next in this literature review, we will present the first generation of experiments (2008-2011), detailing their working principles. Then, we argue that the second wave of experiments (2016-2019) are those in which Hawking correlations were observed. This is to show that, despite a strong theoretical development in those years, the field was mainly driven by experimental research.

\subsection{The Hawking effect in condensed matter systems}
\label{subsec:qft-cms}

The study of analogue gravity started as a research field during the Testimonial Years when the Hawking spectrum was considered in modified gravitational models, and the acoustic metric helped to investigate a variety of semi-classical scenarios of quantum fields on curved space-time. However, the acoustic metric allowed for an expansion of the research. Between 1998 and 2001, the theoretical analysis shifted from QFT on curved space-times to fluid mechanics in condensed matter models with an acoustic horizon \cite{Eltsov_labmodelqft_1998,baldovin_non-gravitational_2000,barcelo_analogue_2001}.

These acoustic fields propagating in a condensed matter system would mimic the quantum fields on curved space-time. Spontaneous emission was expected near the acoustic horizon, and, through the analogy, this would be identified with the Hawking radiation \cite{novello_artificial_2002}. Transposing the problem to a condensed matter system entails designing experiments that would support the claim that the Hawking radiation exists.

Unruh's fluid description of the Painlev\'e-Gullstrand metric enabled the engineering of an effective space-time in condensed matter to investigate the propagation of an acoustic field. This idea was applied first in a superfluid $^3$He~\cite{kopnin_critical_1998,jacobson_effective_1998} and then to an atomic Bose-Einstein condensates~\cite{Garay_BEC_2000}. Likewise, the work on accelerated electrons as thermometers allowed for the use of a classical fluid accelerated over a critical velocity to emulate a curved space-time where a surface wave would propagate \cite{schuetzhold_water_2002}.

At this point, it became clear that systems with some type of horizon could effectively mimic a curved space-time. For instance, a dispersive medium could be engineered such that the electromagnetic field would have a horizon ~\cite{leonhardt_optics_1999,schutzhold_dielectric_2002}, for example, via the phenomenology of slow-light \cite{leonhardt_relativistic_2000}. This example proved itself unfeasible \cite{schuetzhold_slowlight_2003}, but moving media emerged as an alternative to experimentally create an effective space-time with a horizon for the electromagnetic field \cite{Brevik_moving_2001,schutzhold_hawking_2005}.

This short, non-exhaustive list of theoretical work showcasing condensed matter as effective space-times is a testimony of the vibrant activity in the field at the turn of the 21$^{\text{st}}$ century --- an indication of the expansion of the research programme on analogue gravity. We notice the growth of two mutually enhancing, underlying characteristics. First, considerations on the ``Hawking radiation at a black hole horizon'' turned into claims of a ``generalised Hawking effect''. This was seemingly accompanied by philosophical questions on the universality of this effect \cite{unruh_universality_2005}. Second, the viability of analogue models in condensed matter systems was tied to the possibility of realising them experimentally and the reproduction of the effect therein.

The foundation of the analogue gravity programme, then became connected to the understanding of the Hawking effect as a scattering phenomenon, i.e., as a scattering field mode at a horizon and the resulting mixing of quantum field operators yielding emission (as derived by Hawking in 1974 \cite{hawking_black_1974}). 

The expectation value in a field mode of frequency $\omega$ is found by calculating the number operator $\hat{a}_\omega^\dagger \hat{a}_\omega$, which is obtained from the scattering matrix that transforms \textit{in} modes into \textit{out} modes via a Bogoliubov transform~\cite{birrell_quantum_1994,bogoljubov_new_1958}.
If positive and negative (Klein-Gordon) norm modes are non-separable, the \textit{in} vacuum state is not annihilated by the \textit{out} annihilation operator (and vice-versa), and the \textit{out} mode will be populated with field excitations (`particles').

In scattering theory, all the statistics of emission (field amplitudes and correlations) may be obtained from the scattering matrix that connects the incoming field operators with the outgoing ones \cite{bogoljubov_new_1958,macher_blackwhite_2009,recati_bogoliubov_2009}. Therefore, it is reasonable to assume that engineering a scatterer (the stationary background on which the field lives) that implements the same scattering matrix as the event horizon would yield comparable physics (the Hawking effect). In this framework, the Hawking effect describes all scattering phenomena at horizons \cite{macher_blackwhite_2009,recati_bogoliubov_2009,jacquet_influence_2020}.

Typically, vacuum fluctuations of the acoustic field scatter onto at the acoustic horizon~\cite{macher_blackwhite_2009,recati_bogoliubov_2009}, which yields emission by the Hawking effect~\cite{unruh_experimental_1981}.
In modern terminology, Hawking radiation is the wave escaping from the (acoustic) horizon, while the infalling wave is dubbed the `partner radiation'.

The generality of this scattering effect and its identification with `Hawking radiation of black hole horizons' has been at the heart of the epistemological discussion within the physics community since the early 2000s, which can be summarised by the question ``what can we learn from analogue gravity?'' (see e.g.~\cite{parentani_what_2002,visser_analogue_2002,jacobson_qftcshr_2002,volovik_what_2003,unruh_universality_2005}). Despite the lack of consensus in this fundamental question \cite{Field_2021,crowther_what_2018,dardashti_what_2019}, the quest for the Hawking effect has been motivated by the connection with the astrophysical scenario both in theoretical and experimental investigations.

Conceptually, the basis for these investigations is (i) the engineering of an effective space-time and of a scatterer, and (ii) the measurement of outgoing scattering statistics as a function of the incoming state \cite{jacquet_next_2020}.

Note that the state at the input does not need to be the vacuum of the field to enable the observation of the Hawking effect.
It can be, for example, a classical coherent state, a thermal state, or even a squeezed state~\cite{agullo_quantum_2022}.
In the case of a coherent or thermal input state, the emission at the output will not be \textit{spontaneous} (coming from vacuum fluctuations), but it will still be ruled by the scattering matrix of the Hawking effect.
Although always present and broadband, the quantum vacuum is generally very small with respect to the (often monochromatic) coherent component of the input state (see e.g.~\cite[p.~161]{jacquet_negative_2018}).
Therefore, if there is a coherent state at the input, \textit{spontaneous} emission is overpowered by \textit{stimulated} emission --- observing the scattering of a coherent state at the horizon amounts to observing the stimulated Hawking effect.
This is typically what has been observed in experiments to date, starting with demonstrations in fibre optics and an open water channel flow in 2008.

\subsection{The first generation of experiments (2008-2011)}\label{subsec:expprincip}

After decades of theoretical development leading toward measuring the Hawking effect in the laboratory, the first experiments appeared at the end of the 2000s.
These experiments had the clear goal of observing the Hawking effect by designing effective space-times with horizons. Here, we describe the principles of the experiments aimed at observing the Hawking effect: first, how the effective space-times are engineered, and second how the measurements are performed.

\paragraph{Engineering the effective space-time}

All experiments  aiming at observing the Hawking effect at the horizon share common features: 
\begin{enumerate}
    \item The scattering phenomenon is to be studied in a 1+1D space, one spatial plus one temporal dimension. This is possible because of the rotational symmetry of the Schwarzschild space-time.
    \item The physics of interest lies in the regime of linear interactions between linearised excitations, which are, in essence, small amplitude excitations with respect to the background amplitude.\footnote{A typical example of linearised excitations are Bogoliubov excitations in quantum fluids \cite{bogoljubov_new_1958,recati_bogoliubov_2009}.}
    \item These excitations are linearised with respect to a stationary background (the medium) simulating the Schwarzschild space-time in Painlev\'e-Gullstrand coordinates~\cite{visser_acoustic_1998}, more specifically, the one featuring a horizon for the group velocity of excitations.
    \item The field excitations, be they in their vacuum state or not, follow the dispersion relation of the medium~\cite{macher_blackwhite_2009}.
\end{enumerate}

The 2008 experiments in optical fibres~\cite{philbin_fiber-optical_2008} and in the open water channel flow~\cite{rousseaux_observation_2008} exemplify the main two families of analogue systems: those with moving and those with static horizons, respectively (as seen from the laboratory frame where measurements are made).
The latter family is the easiest to visualise: just like in the waterfall metaphor~\cite{unruh_analogue_2007} presented in Fig.~\ref{fig:fish}, the fluid is made to accelerate until its flow velocity exceeds the speed of excitations within, typically the speed of sound in quantum fluids~\cite{recati_bogoliubov_2009}, or the speed of capillary (or gravity) waves in water channel experiments~\cite{schutzhold_gravity_2002}.
The intangible surface thus formed by the moving fluid is static in the laboratory frame.
This horizon separates two regions of the effective space-time, the outside, where the flow velocity of the fluid is lower than the speed of excitations, and the inside of the horizon, where the flow velocity of the fluid is larger than the speed of sound.\footnote{
Depending on the dispersion of excitations in the fluid, negative-norm modes may come from the outside region (as in water waves~\cite{unruh_sonic_1995}) or the inside region (as in quantum fluids~\cite{brout_hawking_1995,corley_hawking_1996,corley_particle_1997}).
This dispersion-induced kinematic feature of fluid-based analogues seems not to affect the Hawking effect~\cite{unruh_sonic_1995,brout_hawking_1995,jacquet_influence_2020}.}

As for the moving horizons, they are realised with a moving modulation of the refractive index of the optical medium~\cite{schutzhold_hawking_2005,nation_analogue_2009}.
Consider the case of light in an optical fibre~\cite{philbin_fiber-optical_2008}: a photon is launched in the fibre just before the modulation of the refractive index, which travels slightly faster than the photon.\footnote{Naturally, this picture also works for vacuum fluctuations of the electromagnetic field inside the fibre~\cite[p.~14]{leonhardt_essential_2010}.}
After some propagation distance, the modulation will catch up with the photon.
Under cross-phase matching, the photon will experience the modulation as a transient increase in the refractive index --- it will be slowed down and trapped.
The front of the modulation separates the outside, in this case, before the modulation where the photon travels faster, from the inside, under the modulation where the photon is slowed down, of the horizon~\cite[p.~43]{jacquet_negative_2018}.
Note that, as seen from the frame co-moving with the modulation, only the photon moves as it slows down and eventually gets trapped by the horizon. But in the laboratory frame, both the horizon and the photon move, albeit at different speeds.

\paragraph{Observing the Hawking effect}

The Hawking effect in all experiments consists in the scattering of paired excitations on either side of the horizon;\footnote{There are other modes because of dispersion, but they do not play a significant role in the present discussion (for more on this issue, see e.g.~\cite{jacquet_influence_2020}).}
the Hawking radiation escapes to the outer region while the partner radiation falls inside the horizon.
The strategy to observe these excitations differ greatly depending on the medium and the type of horizon (moving or static), but, in essence, they may again be separated between fluid-based experiments and optical experiments.

As a typical example of the first type, consider the experiments in water channels, in which the Hawking effect is identified as the partial reflection and transmission of an incoming capillary wave (coherent state) at the horizon.
The Hawking radiation and its partner may thus be observed as amplitude modulations of the water surface in either region.
The acceleration of the water flow is controlled by a physical obstacle while the incoming wave is generated by a wave maker on one end of the channel, outside the horizon~\cite[p.~53]{euve_interactions_2017}.
In 2008, the groups of Germain Rousseaux and Ulf Leonhardt reported observing the scattering of incoming waves into partner waves in Nice, France~\cite{rousseaux_observation_2008}.
This was followed by the observation of the Hawking spectrum in a similar setup by Silke Weinfurtner, William Unruh and collaborators in Vancouver, Canada~\cite{weinfurtner_measurement_2011}.\footnote{Nowadays, it is well established that dispersion in all systems modifies the Hawking spectrum, which is not thermal~\cite{isoard_departing_2020}.}

On the other hand, in optical fibres, moving horizons are realised by coupling intense and ultra-short pulses into the fibre.
Scattering of an incoming wave into Hawking radiation and its partner amounts to partial reflection and transmission through the pulse, respectively.
Due to the Doppler effect (the pulse moves at about $\frac{2}{3}c$ inside the fibre) and dispersion, the reflected light is significantly red-shifted with respect to the incoming wave's frequency. Transmitted light is blue-shifted all the way to the UV, around \SI{220}{\nano\meter} for most commercially available fibres.
In practice, the Hawking effect is observed via these frequency shifts while monitoring the amplitudes at the respective frequencies~\cite{choudhary_efficient_2012}.
In 2008, the groups of Friedrich K\"onig and Leonhardt reported on observing the scattering of an incoming wave into Hawking radiation and measured its spectrum in St. Andrews, UK~\cite{philbin_fiber-optical_2008}.
Because observations in the UV are difficult in optics, direct scattering into the partner mode was only observed later on by the group of Leonhardt in Rehovot, Israel \cite{drori_observation_2019}.

\subsection{The second generation of experiments (2016-2019)}
\label{subsec:2010s}

Over the course of the 2010s, effective 1+1D space-times in other media were demonstrated: in quantum fluids by the groups of Jeff Steinhauer in Haifa, Israel (atomic BEC)~\cite{lahav_realization_2010}, Jacqueline Bloch and Alberto Amo in Paris Saclay, France (superfluid of microcavity polaritons)~\cite{nguyen_acoustic_2015} and Peter Skyba in Ko\v sice, Slovakia (superfluid $^3$He-\textit{B})~\cite{clovecko_magnonic_2019}, as well as in bulk optical crystals by the group of Daniele Faccio in Como, Italy~\cite{faccio_analogue_2010}.

The latter led to a chain of unconvincing results: In 2011, Faccio's group claimed to have observed spontaneous emission from a moving horizon~\cite{belgiorno_hawking_2010}, but the results were deemed inconclusive~\cite{ belgiorno_quantum_2010,belgiorno_dielectric_2011,rubino_experimental_2011}\textcolor{red}{\cite{liberati_quantum_2012,unruh_hawking_2012}}.
Since then, analytical~\cite{finazzi_quantum_2013,finazzi_spontaneous_2014,jacquet_quantum_2015} and numerical~\cite{bermudez_hawking_2016} calculations have established that the Hawking effect does not occur in the configuration of~\cite{belgiorno_hawking_2010}.

Beyond the observation of wave scattering at the horizon, analogue gravity enables the observation of correlations across the horizon~\cite{carusotto_numerical_2008}.
This is a marked difference with black holes, from within which no signaling is possible.
The second half of the decade saw the rise of a new generation of experiments specifically aimed at observing correlations across the horizon.

\paragraph{Analysis of the experiments' principles}

In water channels, thermal fluctuations largely overpower quantum fluctuations, thus preventing the observation of spontaneous emission. 
However, the signal-to-noise ratio in quantum fluids such as BECs or superfluids of microcavity polaritons should allow for such measurements.

In 2016, the group of Germain Rousseaux and Renaud Parentani in Poitiers, France, monitored classical noise on the water interface surrounding horizons and measured Hawking correlations in this system \cite{euve_observation_2016}.
Similar measurements were reported by Steinhauer that same year~\cite{steinhauer_observation_2016}, leading to the observation of Hawking correlations in an atomic BEC in 2019~\cite{munoz_de_nova_observation_2019}.
To date, there is still a debate as to whether spontaneous emission from the quantum vacuum has been observed in atomic experiments~\cite{michel_phonon_2016,leonhardt_questioning_2018,steinhauer_comment_2018,coutant_low-frequency_2018}.
One way of discriminating the thermal amplification from the quantum noise would be measuring the entangled content of observed Hawking correlations \cite{robertson_assessing_2017,coutant_low-frequency_2018-1,isoard_bipartite_2021}.
Even if spontaneous emission has not yet been confirmed, the correlations evidence pair-wise scattering at the horizon by the Hawking effect, and so this aspect of quantum field theory on curved space-time has been verified in the laboratory.

Measuring these correlations requires a different set of tools and methods, and, naturally, the statistics obtained from them are also different. Therefore, these experiments reported in \cite{euve_observation_2016,steinhauer_observation_2016,munoz_de_nova_observation_2019} mark the advent of the second generation of experiments.

\paragraph{A transition?}

Theoretical work in analogue gravity also expanded in the meantime.
The advent of experiments in 2008 prompted a large number of theoretical works that analysed their results and proposed experimental metrics to measure and gain insight into the Hawking effect in condensed matter systems. Listing or reviewing all of these reports goes beyond the scope of this paper.\footnote{Google Scholar returns over 1,500 results for the search ``analogue gravity'' between Jan. 1$^{\text{st}}$, 2008 and Sept. 21$^{\text{st}}$, 2022.} Suffices to say that a significant subset of the theoretical production --- including the examples listed above \cite{michel_phonon_2016,leonhardt_questioning_2018,steinhauer_comment_2018,coutant_low-frequency_2018} --- seems to be driven by experimental data or proposed to improve experimental performance.

Together with new proposals for experiments in other condensed matter systems (see, for example, \cite{Marino_fluidlight_2008,Nation_squidHR_2009,Horstmann_ionHR_2010}), this shift towards an experimental drive of the field opens the path to exploit the Hawking effect as a benchmark for other phenomenology studies in condensed matter systems and signals a new phase of research for the analogue gravity programme.

\section{Autonomy of the research programme}
\label{sec:autonomous}

The experimental observation of the Hawking effect in different condensed matter systems between 2008 and 2019~\cite{philbin_fiber-optical_2008,weinfurtner_measurement_2011,euve_observation_2016,munoz_de_nova_observation_2019} marked the culmination of the historical phase during which analogue gravity aimed at confirming the existence of that effect of quantum field theory on curved space-times and establishing the potential of condensed matter systems for such investigations~\cite{jacquet_next_2020,Field_2021}.
We will now show that contemporary research in the field has three main dynamics that go beyond the original purpose of analogue gravity: studying new field-theoretic phenomena (other than the Hawking effect), improving our understanding of condensed matter systems, and predicting new field-theoretic effects. While the first two dynamics are well established, the last one is only emerging.

We argue that this departure from the initial purpose of analogue gravity experiments signals a gradual autonomisation of the research programme from the Hawking effect.
This does not mean that observing the Hawking effect lost relevance but that the intrinsic value of the investigation becomes independent from the question of whether it confirms the existence of Hawking radiation.

\subsection{Looking beyond the Hawking effect}\label{subsec:beyondHE}

Considerations beyond the Hawking effect appeared in theoretical research on analogue gravity already in the early 1990s, for example, in the work of Matt Visser. Visser wrote a paper that first appeared in a preprint on the website ArXiv in 1993.
Its first version titled \textit{Acoustic propagation in fluids: an unexpected example of Lorentzian geometry} \cite{visser_acoustic_1993}
was essentially an independent rediscovery of the equivalence between the wave equations of the acoustic field in an inhomogeneous fluid flow and the scalar field on a curved space-time\footnote{Much like the early work of White~\cite{white_acoustic_1973} and Anderson and Spiegel~\cite{anderson_radiative_1975}.} but without the application of quantum field theory. 
Visser learned from Ted Jacobson and John Friedman about Unruh's 1981 paper and added a note post-submission acknowledging him.
Visser revisited his results and expanded in a second version of the paper, re-titled \textit{Acoustic black holes: horizons, ergospheres and Hawking radiation} \cite{visser_acoustic_1998}.
In it, Visser generalises Unruh's analogy, discussing other geometries, like 2+1D vortex flows that feature additional surfaces to the event horizon that opened the way to the study of field effects on rotating black-hole (so-called Kerr) geometries in their equatorial slice. These ideas were soon adapted to cosmological models~\cite{Bassett_cosmo_2000, volovik_vacuum_2001}\footnote{These early proposals were quickly followed by a number of more realistic calculations, notably in atomic BECs proposing operational means to measure effects like inflation and cosmological particle pair creation, see for example~\cite{fischer_cosmo_2003,fischer_particles_2004}.}.

The review by Barcel\'o \textit{et al.}~\cite{barcelo_analogue_2011} summarises the theoretical activity in the field until 2010, when it was last updated.
The topics considered therein reach far beyond the Hawking effect in Schwarzschild geometries.
The scope of the use of the analogy had broadened (at least as far as theory was concerned), tackling other issues in astrophysical and cosmological scenarios.
This indicates that the autonomisation of the research programme was already in motion in the early 2000s~\cite{visser_analogue_2002}.
So, when experimental research began in the late 2000s, this expansion on the analogy allowed for a greater exploration of condensed matter systems. In many instances, the table has turned: the analogy to the astrophysical/cosmological scenario is now helping to improve our knowledge of condensed matter systems.

Here we will present a narrower literature review than Barcel\'o's, focusing on the new (third) generation of experiments to show how the present dynamics result from the experiment-driven autonomisation of the mid-2000s.
Naturally, this selection is not representative of the whole, but it is a significant subset that investigates analogue gravity under expectations that diverge from the historical purposes presented so far.

There are, broadly speaking, three main families of such experiments: those on rotating geometries to study fields around Kerr black holes, those in temporally evolving geometries to study fields in various cosmological scenarios, and those studying field equations with dynamical metrics.
Because of its novelty, we will not discuss the latter family but recommend the 2021 experimental reports~\cite{kolobov_observation_2021,patrick_backreaction_2021} and the theoretical paper~\cite{fabbri_ramp-up_2021} for a recent discussion of this topic. In our view, this new trend only reinforces our conclusions.

\paragraph{Field effects in rotating geometries}

Rotating fluid flows such as vortices mimic the Kerr geometry, i.e., a rotating black hole~\textcolor{red}{\cite{schutzhold_gravity_2002, richartz_rotating_2015}}.
Because they feature other surfaces of interest outside the horizon, like the ergosurface or the light-ring~\cite{visser_acoustic_1998}, such geometries enable the study of numerous interesting effects besides the Hawking effect (which has neither been studied much nor observed therein).\footnote{More on these studies may be found in the references listed in~\cite{patrick_analogy_2020}.}

Silke Weinfurtner's group has shown rotational superradiance in experiments with water waves in 2017~\cite{torres_rotational_2017}, with similar measurements reported by the group of Faccio in the vortex flow of a fluid of light in 2022~\cite{braidotti_measurement_2022}.\footnote{The title of that paper is a misnomer, the observed wave effect is rotational superradiance~\cite{zeldovich_generation_1971,richartz_generalized_2009}, which is different from the Penrose effect~\cite{brito_superradiance_2020} (the reflection of particles on the ergosurface that is accompanied by an increase in their momentum associated with a decrease in the momentum of the rotating black hole~\cite{solnyshkov_quantum_2019}).}
Rotational (or Zel'dovich) superradiance is the reflection and amplification of incident waves by a rotating body~\cite{zeldovich_generation_1971,richartz_generalized_2009}.
This is comparable to the Hawking effect in that amplification is also due to the mixing of positive- and negative-norm modes of the field at a specific surface of the effective space-time (in this case, the ergosurface instead of the horizon~\cite{dolan_scattering_2013}).

Weinfurtner's group also reported on the oscillations of the light ring of the water vortex in 2020~\cite{torres_quasinormal_2020}, which is due to the excitation of quasi-normal mode.
These are the relaxation mechanism of a perturbed system as it falls back to its equilibrium state~\cite{berti_quasinormal_2004}.
A water vortex like the draining bathtub flow of~\cite{torres_quasinormal_2020} is never effectively at equilibrium, and so it spontaneously relaxes by populating the light-ring modes. Because these are the lowest energy modes capable of traveling across the entire system, they represent the most efficient way to radiate extra energy away in vortex flows.

Both effects could have been anticipated from Visser's 1998 paper (see~\cite{cardoso_stability_2009}) and were theoretically derived and studied long before they were observed (see e.g.~\cite{schutzhold_gravity_2002,berti_quasinormal_2004,dolan_scattering_2013,dolan_resonances_2012}).
Thus, these experiments contributed to settling the ongoing transition of the field towards a generalised study of scattering effects on effectively curved space-times --- a new role for analogue gravity, as identified by Fields \cite{Field_2021}.

\paragraph{Field effects in temporally evolving geometries}

Any modification of the space-time geometry will lead to the mixing of waves and emissions from the vacuum.
In the case of black holes, this is due to a spatial inhomogeneity, like the phenomenon of rotational superradiance discussed above.
However, the emission can also occur when the geometry evolves in time, for example, when the space-time is expanding~\cite{schrodinger_proper_1939,parker_particle_1968}.

Emission in an expanding effective space-time may be observed in the laboratory with experiments in which the size of the medium is made to grow. Pair creation was observed in an expanding trap of ultra-cold ions by the groups of Tobias Schaetz and Ralf Sch\"utzhold in 2019~\cite{wittemer_phonon_2019}.
This scattering effect is tied to the rate of spatial expansion of the effective space-time, the so-called Hubble friction parameter, which was measured in an expanding trap of ultra-cold atoms~\cite{eckel_rapidly_2018} by the group of Ian Spielman and Gretchen Campbell in 2022~\cite{banik_accurate_2022}.

A change in the size of the medium modifies the effective interactions within, which leaves behind oscillations in the spectrum of fundamental excitations called Sakharov oscillations~\cite{sakharov_initial_1966}.
This was observed in a variety of atomic media by the groups of Cheng Chin in 2013~\cite{hung_cosmology_2013}, Chen-Lung Hung in 2021~\cite{chen_observation_2021}, Alberto Bramati and Quentin Glorieux in 2022~\cite{steinhauer_analogue_2022} and Markus Oberthaler in 2022~\cite{viermann_quantum_2022}.
Similarly, a temporal modulation of the density of the medium yields emission in a process akin to the dynamical Casimir effect, as was demonstrated in an atomic BEC by the group of Chris Westbrook in 2012~\cite{jaskula_acoustic_2012}.\footnote{Similar methods to~\cite{jaskula_acoustic_2012} applied in Josephson metamaterials allow to measure the dynamical Casimir effect, as reported in 2011~\cite{wilson_observation_2011}. In fact, the results of~\cite{jaskula_acoustic_2012} were initially interpreted as an observation of the dynamical Casimir effect, but it is now theoretically established that the experiment rather reproduces the dynamics of preheating~\cite[p.~2]{busch_quantum_2014}.}

At this point, it is clear that the phenomena observed are not attached solely to the astrophysical or cosmological scenario since it is a generic result of quantum field theory on curved spaces. Although these new effects were theoretically tied to the creation of effectively curved space-times in the laboratory, the experimental measurements make them intrinsically a condensed-matter result --- a novel result. 
In this light, we now discuss how the analogue gravity programme impacts research on condensed matter.

\subsection{Analogue gravity for condensed matter systems}

Analogue gravity has been driving experimental developments in condensed matter ever since the advent of the first experiments in 2008.
In some cases, the scope goes beyond the observation of field effects initially predicted in relativistic contexts.
For example, the ease-handling of optical experiments enables one to finely tune the amount of frequency shifting at the horizon \cite{choudhary_efficient_2012}. This mechanism for frequency shifting has played a key role in a variety of nonlinear phenomena in this system \cite{frisquet_optical_2016}.
Because of this, methods for optical horizons~\cite{demircan_controlling_2011} have been used in fundamental as well as applied experiments since the 2010s~\cite{leonhardt_cosmology_2015}, from the engineering of pulse-wave~\cite{hill_evolution_2009,bose_experimental_2015,ciret_observation_2016} and pulse-pulse~\cite{tartara_soliton_2015,kanakis_enabling_2016} interactions to the generation of coherent and incoherent supercontinua~\cite{wang_optical_2015,bose_fiber_2019,yang_observation_2021} and even optical rogue waves~\cite{frisquet_optical_2016}.

More recently, the observation of the relaxation of a vortex by quasi-normal modes in 2020~\cite{torres_quasinormal_2020} opened a new avenue to apply methods from relativity to condensed matter systems.
In astrophysics, the QNM spectrum does not depend on the characteristics of the object that perturbed the black hole, but only on the black hole's mass and angular momentum~\cite{abbott_observation_2016}.
According to general relativity, together with the electric charge, these are the only three parameters needed to fully describe the black hole.
Therefore, observing QNMs allows for the full characterisation of the perturbed black hole by a process known as black-hole spectroscopy~\cite{echeverria_bhs_1989}.
The same goes for vortex flows, they may be entirely characterised by the spectroscopic study of their QNMs~\cite{torres_waves_2018}.
This means that vortices in fluids and superfluids alike may be characterised by the study of their light ring spectrum, which may be dubbed analogue black hole spectroscopy~\cite{torres_analogue_2019}.

These two cases illustrate how methods from analogue gravity may be used beyond the field of analogue gravity: theory on astrophysics and cosmology is proving to be a guide to further knowledge in condensed matter.
These methods call on concepts from the relativistic context that happen to apply in the condensed matter as well, whereby the analogy is somewhat reversed: quantum field theories acknowledging the comparison to the astrophysical system to unveil hidden features of condensed matter systems.
However, although the theory of finding new properties of condensed matter systems comes from astrophysical scenarios, it does not depend on the empirical validation of the astrophysical phenomena to verify the new-found condensed-matter effects. The reverse analogy is not intended to provide evidence in this case but to further the theory in condensed matter. This is the same purpose we have identified in the Testimonial Years for the astrophysical scenario.

\subsection{A new phase}\label{subsec:newphase}
Finally, we briefly touch on the potential of analogue gravity to predict new effects in quantum field theory.
This poses the question about the potential of condensed matter systems as analogue quantum simulators of field theories on curved space-times~\cite{barcelo_analogue_2011,steinhauer_viewpoint_2012,weinfurtner_quantum_2019,liberati_information_2019,mccormick_uncovering_2022}.
Indeed, their capability to make predictions about universal phenomena is known to be limited~\cite{dardashti_what_2019,crowther_what_2018,evans_limits_2020,Field_2021} and it may be that a new effect discovered in this context would not exist for field theories in the relativistic context.
Nevertheless, the emergence of such new proposals~\cite{leonhardt_case_2020,jacquet_quantum_2022} defines a new trend in the research programme.

In this section, we have reviewed the literature and found that contemporary research in the field has moved beyond the observation of the Hawking effect.
There is a new generation of experiments that aims at studying phenomenology associated with quantum fields in rotating black-hole geometries or cosmological scenarios (such as inflation or preheating).
Furthermore, the analogy is being reversed to use concepts from the relativistic context and methods of analogue gravity to study the condensed matter systems.
It may also be that analogue gravity will be proved useful to discover new field-theoretic effects, although only time will settle the matter.
These new trends in the field have grown in strength at different paces as theoretical work is becoming more and more driven by experiments.
It inspires a reassessment of the role and application of analogue gravity in quantum field theory research.

We also remark that there still are experiments on the Hawking effect in 1+1D geometries: building on the second generation of experiments identified in Subsection~\ref{subsec:2010s}, these aim at testing the effect in different configurations to study its robustness~\cite{euve_scattering_2020,kolobov_observation_2021} and, possibly, its interplay with other phenomena like the excitation of quasi-normal modes, for example~\cite{nguyen_acoustic_2015,jacquet_quantum_2022}.
These are not completely aimed at \textit{confirming} the effect, but also they are \textit{used} as a benchmark for other phenomenology.

The autonomisation of the research programme refers to the fact that the analogy acquired new purposes other than the confirmation of astrophysical predictions. The philosophical question of whether an observation in the laboratory provides empirical evidence for astrophysical phenomena remains open, but the research on analogue gravity does not \textit{need} a positive answer anymore to justify its existence. The reverse analogy justifies itself independently from the discussion on the ontology of the astrophysical theory. This turn of events is not actually something new. It was foreseen in Paul David's words about Fulling's paper as early as 1975 (see Subsection \ref{subsec:mirror}). ``If these conclusions
also apply (...), then the whole character of discussion about gravitational collapse seems to be changed. // The consequences of Hawking’s result for elementary particle physics are no less
drastic.'' \cite[p.~616]{davies.1975} Of both of these promises, it was the discussion on Hawking radiation that guided the analogies and experimental proposals. But the research programme has expanded and broadened its objectives  beyond the observation of the Hawking effect.

The new roles for analogue gravity we noticed in this current trend can be attributed to the expansion of the research programme. An increase in the number of people working in the field allows the investigation to depart from the mainstream. A similar situation occurred in the field of general relativity itself. Over thirty years, starting from the mid-1920s, the field was generally perceived as sub-disciplinary. Blum \textit{et. al} argued that the renaissance of general relativity after this period of low water-mark happened in parts due to the establishment of a long post-doctoral education tradition in the 1950s \cite{blum_renaissance_2016}. Today as well, the analogue gravity research programme seems to be acquiring enough independence from the gravitation field of research.

\section{Conclusion}

The quantum investigation of a black-hole environment, resulting in the prediction that black holes could radiate and eventually evaporate, brought more than new theoretical developments to the table. It allowed for specific comparison between it and different physical systems, transposing the astrophysical scenario to accessible models that could, in principle, be reproduced on Earth. At each stage of research, the use of analogies had different purposes. Taking this into consideration, we divided the historical eras into three plus one periods of research on the Hawking effect through analogue models and classified them according to the objectives for analogue reasoning. 

In the Formative years (pre-1974), Yakob Zel'dovich compared the classical gravitational theory governing black holes to a resonator system in which quantum effects revealed essential features. Most notably, in this analogue system, it was expected that the resonator would reproduce an effect similar to the Penrose Process. The analogy indicated that the Penrose mechanism to extract energy from a rotating black hole was plausible, but the analogy presented odd results perceived as probable flaws. Zel'dovich realised that the amplification of waves in a resonator, the analogue phenomenon to the Penrose Process, would lead to wave generation near the horizon, which might not translate well to the astrophysical scenario. Around the same time, William Unruh, during a lecture on the subject, came up with a sonic-fluid analogue with pedagogical purposes: a fish falling in a waterfall flowing with a velocity greater than the speed of sound. This fish-in-the-waterfall analogy was used briefly as a visualisation tool---an artifice to understand how light would be trapped inside the event horizon of a black hole. This analogy could have been completely forgotten if not for the later development when Unruh expanded the calculations to prove the analogy was mathematically sound. However, the most transformative analogy came from Jacob Bekenstein. He compared black-hole physics with the theory of thermodynamics, which led to a new understanding of black holes. Although they were different in nature, all of those analogies helped to further our understanding of black holes, even though they could present false conclusions. Zel'dovich's analogy and Bekenstein's theory meant black holes could radiate, an odd result when the general theory clearly stated that nothing could escape the event horizon.

The situation changed in 1974, when Hawking applied quantum field theory in the vicinity of a black hole and concluded that, indeed, they should radiate. The extreme curvature around the horizon meant more particles would be created and sent to infinity than annihilated or falling into the hole. Hawking's results indicated that there could be more to the thermodynamical analogy than previously thought. It lost the status of analogy altogether, and black holes were widely accepted to be thermodynamical objects, the closest to a black body in nature. It was the beginning of the Testimonial Era (1974-2008) when analogies and effective theories were proposed as evidence that Hawking radiation was plausible. Unlike Zel'dovich's analogy, however, the ones proposed in this period did not appear to lead to false conclusions, being a more credible representation of the astrophysical scenario. Several viable experiments appeared in the hopes of replicating the phenomenon predicted in the astrophysical system in the laboratory. The analogies and effective theories were justified by the fact that they could reproduce this Hawking-like phenomenon. 

Over the years, the analogue phenomenon became known as the Hawking effect, a scattering occurrence near a horizon-type region to which the Hawking radiation was only one example. The experimental search for the Hawking effect became the focus of the analogical reasoning in 2008, with a concrete proposal of an experiment using fibre optics. In this case, the role of the analogy was to provide theoretical confirmation to the astrophysical phenomenon.

Given the relevance of the topic, we hope that this study will be useful for historians and philosophers to assess the use of analogue gravity. The topic is rich and straightforwardly aligns with philosophical discussions. For example, it is interesting to analyse how the three eras can be interpreted in Paul Bartha's classification for prospects for analogue confirmation \cite{bartha_2022}. The analogies from the Formative Years seem to closely follow what Bartha called \textit{explanatory} analogical arguments in a \textit{weak} logical structure, in the sense that those analogies were used to support the explanatory hypothesis to explain known features, rendering the comparison worthwhile with potential for generalisation. The Testimonial and Empirical Years feature \textit{predictive} analogies with seemingly \textit{intermediate} and \textit{strong} inductive supports, respectively.

In the spirit of bringing contemporary topics in physics to their historical context, we propose that the modern era of research configures a different phase, the Autonomous Phase, characterised by the independence of the research on analogue models to the original questions about the astrophysical counterpart. Because it is a conjecture, we intentionally broke the pattern we adopted by calling this a \textit{Phase} instead of \textit{Years} to set it apart from the established historical periods. To support our claim, we offered a literature review on the new trends in the analogue gravity research programme, showing that the analogies to the astrophysical scenario are the basis for new phenomenological discoveries in the field of condensed matter in a move that can be considered a reverse-analogy. When before the possibility of testing the thermodynamic theory of black holes was the main propeller to build experiments, now the benefits are reciprocal to the point where the justification for studying analogue gravity would survive even without the belief that it could provide empirical evidence of the existence of Hawking radiation.

\bmhead{Acknowledgments}
While brainstorming this manuscript, trying to connect history and modern physics, we have benefited enormously from insightful discussions with Grace Fields. We thank her wholeheartedly for those conversations and for carefully reading this manuscript and providing feedback that certainly helped us improve our work. We also thank William Unruh for his readiness to answer a few questions that had appeared during the research process and for allowing us to use his original drawing of a fish in a waterfall in this recounting.

\clearpage

%\bibliography{bibliography.bib}

\end{document}